\documentclass[preprint]{ptptex}
\usepackage{graphicx}

\newcommand{\vev}[1]{{\langle #1 \rangle}}
\newcommand{\abs}[1]{{\left| #1 \right|}}

\newcommand{\GeV}{\mbox{~GeV}}

\newcommand{\ie}{{\it i.e.}}
\newcommand{\eqn}[1]{&#1&}
\notypesetlogo

\title{Non-perturbative Corrections to Particle Production 
 from Coherent Oscillation}

\author{%
Takehiko \textsc{Asaka}$^1$ 
and
Hiroaki \textsc{Nagao}$^2$
}

\inst{%
$^1$Department of Physics, 
Niigata University, Niigata 950-2181, Japan
\\
$^2$Graduate School of Science and Technology, 
Niigata University, Niigata 950-2181, Japan
}

\abst{ 
  We investigate particle production from coherent oscillation by
  using the method based on the Bogolyubov transformation.
  Especially, we study the case when the amplitude of the oscillation
  and also the coupling constants with the oscillating field
  are small in order to avoid the non-perturbative corrections from
  the broad parametric resonance.  We derive the expressions for 
  the distribution functions and the number densities of produced particles
  at the leading order of coupling constant.  
  It is, however, found that these results fail to describe the exact
  particle production eventually due to the
  non-perturbative effects even if the coupling constants are small.
  We then introduce a simple method
  to handle with such corrections, \ie, the time averaging method.
  It is shown that this method successfully provides the evolution of
  the occupation numbers of the growing mode.
  Further, we point out that the approximate results by this method
  satisfy the exact scaling properties coming from the periodicity
  of the coherent oscillation.
}
\begin{document}
\maketitle
%
%
\section{Introduction}
Coherent oscillation of scalar field plays an important role to
describe various phenomena in particle physics and particle cosmology.  
One of the most
important examples is the so-called slow-roll model of the
inflationary universe.~\cite{SRINF,LindeBook} A scalar field, called
as inflaton, is initially displaced from the potential minimum and its
vacuum energy leads to the de-Sitter expansion of the universe.  This
class of models elegantly solves the flatness and horizon problems of
the standard Big Bang cosmology.  Moreover, it can give an origin of
the density fluctuations which is strongly supported from the recent
measurements of the cosmic microwave background
radiation.~\cite{Larson:2010gs} After the inflation ends, the
inflaton starts to cause coherent oscillation around its potential
minimum.  

The energy of coherent oscillation is diluted due to the expansion of
the universe as well as the energy transfer to particles though
interaction of the oscillating field.  Produced particles are then
thermalized and the hot universe can be realized.  The whole of these
processes is called as the reheating.  In particular, the reheating of
the slow-roll inflation gives an initial condition of the standard
Big Bang cosmology.  
Therefore, the reheating process is crucial for understanding
the very early universe.

In this paper, we focus on the first stage of the
reheating, \ie, particle production from coherent oscillation.
This process has been widely discussed based on that
coherent oscillation is considered as non-relativistic scalar
particles,~\cite{NR_particle} \ and particles are produced through
their decay and/or scattering processes.~\cite{Albrecht:1982mp} \ In
this case the number of parent non-relativistic particles is given by
the energy of coherent oscillation divided by the mass of the
oscillating field, and the number of produced particles is
determined by it.  On the other hand, it has been pointed out that,
when the coupling with produced particles and also the amplitude of
the oscillation become large, the non-perturbative effect becomes
significant in the early stage of particle
production~\cite{Traschen:1990sw,Kofman:1994rk,Boyanovsky:1994me,Yoshimura:1995gc}. \
This process is called as the preheating.~\cite{Kofman:1994rk} \
Especially, the explosive production of bosonic degrees of freedom can
happen due to the broad parametric resonance effect.  
 On the other hand, the fermion
production at the preheating has been also
investigated.~\cite{Dolgov:1989us,Baacke:1998di,Greene:1998nh}

The purpose of this paper is to investigate the production of scalars
and fermions from coherent oscillation, especially when the coupling
constants of oscillating field are very small to avoid the effect of
the broad parametric resonance.  For this purpose, we apply the method
based on the Bogolyubov
transformations.~\cite{Bogolyubov:1958se,Birrell} In this
case, the equation of motion for the mode functions of produced
particles in the presence of coherent oscillation is solved and the
growth of the mode functions are then interpreted as the production of
particles.  First, we will present the analytical formulae for the
distribution functions and the number densities of 
produced particles by using the perturbative expansion of the coupling constants.
We will also discuss the conditions under which the
perturbative results are justified.  Indeed, it will be shown that the
leading-order results collapse in the end.  

This is a signal that the non-perturbative effect becomes important 
even if the coupling constant is sufficiently small.
Such a correction is crucial for describing the statistical properties of produced
particles, namely the effects of the Bose condensation for the scalar
production and the Pauli blocking for the fermion production.  In
order to handle the annoying non-perturbative effects we will present
the time averaging method, which is familiar in the nonlinear
dynamical system.~\cite{NAYFEH} 
It will be demonstrated that this method is powerful
to extract the characteristic evolution of the occupation number 
for the growing mode, \ie, the exponential growth for scalar production
or the oscillation between 0 to 1 for fermion production.
Furthermore, we will show that the
results by the time averaging method obey the exact scaling property,
which is obtained from the periodicity of the the equation of motion.~\cite{Mostepanenko:1974}  
This gives a justification for the use of the time averaging method.  
Throughout the present analysis we neglect the expansion of the universe for
simplicity.

The rest of this paper is organized as follows.  
In Sec.~\ref{sec:Framework} we explain the model in this analysis. 
We perform in Sec.~\ref{sec:Perturbative_Estimate} the perturbative estimation of
the yields when the amplitude of the coherent oscillation is
sufficiently small.  The importance of non-perturbative effects in
particle production is addressed in Sec.~\ref{sec:Time_Ave}.  We
present the time averaging method to deal with such effects, and try
to figure out the statistical properties of produced particles.
Finally, the last section is devoted to conclusion.
We also add Appendix~\ref{sec:AP1} to explain the perturbative
estimation of the number density.
\section{Framework}
\label{sec:Framework}
To begin with, let us explain the framework of this analysis.
We shall study the production of real scalar field $\chi$ and
Dirac fermion $\psi$ from the coherently oscillating $\phi$
by using Lagrangian
\begin{eqnarray}
    \label{eq:L}
    {\cal L} \eqn{=}
    \frac{1}{2} \, (\partial_{\mu}\phi) (\partial^{\mu}\phi)
    - V(\phi)
    +
    \frac{1}{2} \, (\partial_{\mu}\chi) (\partial^{\mu}\chi) 
    - 
    \frac{1}{2} \, g_S^2 \, \phi^2  \, \chi^2  
    + i \, \overline{ \psi } \gamma_\mu \partial^\mu \psi
    - g_F \, \phi \, \overline{ \psi} \, \psi \,,
\end{eqnarray}
where $g_S$ and $g_F$ are coupling constants.  For definiteness,
we take here the potential for the real scalar field $\phi$ as
\begin{eqnarray}
    V(\phi) = \frac{1}{2} \, m_{\phi}^2 \, 
    \Bigl(\phi -\langle \phi \rangle \Bigr)^2 \,,
\end{eqnarray}
where $m_\phi$ and $\vev{\phi}$ are mass and vacuum expectation value
(vev) of $\phi$, respectively, and they are taken to be real and
positive.  At the potential minimum $\chi$ and $\psi$ receive masses
as $m_\chi = g_S \vev{\phi}$ and $m_\psi = g_F \vev{\phi}$.  
From now on the field $\phi$ is assumed to oscillate coherently 
around $\vev{\phi}$ with an amplitude $\Phi$
\begin{eqnarray}
  \label{eq:PHI_T}
  \phi (t) = \vev{\phi} + \Phi \, \cos ( m_\phi t) \,,
\end{eqnarray}
and it is treated as a classical background field.  Notice that we
neglect the expansion of the universe throughout the present
analysis.

Particle production from the $\phi$ oscillation is usually discussed
as follows: The coherent oscillation is considered as 
non-relativistic particles~\cite{NR_particle}.  
In this case, the energy density of
the $\phi$ oscillation is $\rho_\phi = \dot \phi{}^2/{2} +
V(\phi) = m_\phi^2 \, \Phi^2 /2$ (here and hereafter the dot denotes a
derivative with respect to time),  and the number
density of $\phi$ is estimated as $n_\phi = \rho_\phi /m_\phi = m_\phi
\, \Phi^2 /2$.  Decays of $\phi$, $\phi \to \chi + \chi$ and $\phi \to
\psi + \overline \psi$, are important processes to produce $\chi$ and
$\psi$.  The partial rates of these processes are found from
Eq.~(\ref{eq:L}) as
\begin{eqnarray}
  \Gamma_{\phi \to \chi+ \chi} 
  \eqn{=}
  \frac{ g_S^2 \, \beta_\chi }{8 \pi} \, \frac{m_\chi^2}{m_\phi} \,,
  \\
  \Gamma_{\phi \to \psi+ \overline \psi} 
  \eqn{=}
  \frac{ g_F^2 \, \beta_\psi^3 }{8 \pi} \, m_\phi \,,
\end{eqnarray}
where $\beta_{\chi, \psi} = \sqrt{ 1 - 4 m_{\chi,\psi}^2/m_\phi^2}$.
The number densities of $\chi$ and $\psi$ from the decays of
$\phi$ are then estimated as
\begin{eqnarray}
  \label{eq:Nchi_naive}
  n_\chi(t) \eqn{=} 2 \, \Gamma_{\phi \to \chi + \chi} \, 
  n_\phi \, t
  = \frac{g_S^2 \, \beta_\chi \, \Phi^2 \, m_\chi^2}{8 \pi} \, t 
  \,,
  \\
  \label{eq:Npsi_naive}
  n_\psi(t) \eqn{=} 2 \, \Gamma_{\phi \to \psi + \overline{\psi}} \,
  n_\phi \, t
  = \frac{g_F^2 \, \beta_\psi^3 \, \Phi^2 \, m_\phi^2}{8 \pi} \, t \,.
\end{eqnarray}
It is seen that $n_{\chi,\psi}$ are proportional to $t$
by neglecting the decrease of $n_\phi$, and that they are
induced at the order of coupling squared.

Further, the scattering processes of $\phi$'s are another sources of
particle production.  The number density of $\chi$ due to the process
$\phi + \phi \to \chi + \chi$ is estimated as $n_\chi (t) = \langle
\sigma v_\phi \rangle \, n_\phi^2 \, t$, where $v_\phi$ is the
relative velocity of $\phi$'s and $\langle \sigma v_\phi \rangle$ is
the invariant scattering rate which is given by $\langle \sigma v_\phi
\rangle = g_S^4 \, \beta_{\chi}^s/( 32 \, \pi \, m_\phi^2 )$
with $\beta_{\chi}^s=\sqrt{1-m_\chi^2/m_\phi^2}$
in the non-relativistic limit $v_\phi \to 0$.  We then find that
\begin{eqnarray}
  \label{eq:Nchi_naive_g4}
    n_\chi (t) = \frac{ g_S^4 \, \beta_{\chi}^s \,
      \Phi^4 }{ 128 \, \pi } \, t \,,
\end{eqnarray}
which is again proportional to $t$, while it is induced at the fourth
order of coupling constant.  
Thus, the production via scattering can be neglected 
as long as the oscillation amplitude is sufficiently small,
say $\Phi \ll \vev{\phi}$.
It should be noted that the
scattering rate of $\phi + \phi \to \psi + \overline{\psi}$ vanishes
in the non-relativistic limit, and hence the production of $\psi$ via
scattering is less significant.

It has been discussed in the literature that particle production from
the coherent oscillation is more involved than the above naive
treatment.  In the following, we study the production of $\chi$ and
$\psi$ by using the method based on the
Bogolyubov transformation~\cite{Bogolyubov:1958se,Birrell}.  Especially, we
concentrate on the case in which the coupling constants $g_S$ and
$g_F$ are very small, say $g_{S,F} \ll m_\phi/\Phi$ , in order to
avoid the non-perturbative effect due to the broad parametric
resonance.  We perform both analytical and numerical
estimations of the yields, and find the validity of the naive argument
of particle production.

\section{Perturbative Estimate of Yields}
\label{sec:Perturbative_Estimate}
We are now at the position to derive the analytical expressions for 
the yields of $\chi$ and $\psi$ at the leading order of the 
coupling constant $g_S$ or $g_F$.  
Hereafter, we identify the masses $m_\chi$ and $m_\psi$ as
parameters being independent on $g_S$ and $g_F$, although they are
${\cal O}(g_S)$ and ${\cal O}(g_F)$ quantities.  Moreover, we assume that
the amplitude of coherent oscillation is small as $\Phi \ll
\vev{\phi}$ in order to avoid the production from 
the scattering processes.%
\footnote{ The case of the large amplitude $\Phi \gg \vev{\phi}$ will be discussed in elsewhere.~\cite{AN} }

\subsection{Production of scalar}
Let us first consider the production of the scalar $\chi$.
In the presence of coherent oscillation in Eq.~(\ref{eq:PHI_T}) the
equation of motion for $\chi$ is given by
\begin{eqnarray}
    \label{eq:EOM_CHI}
    \left[ \square - M_\chi^2 (t) \right] 
    \chi (t, \vec x) = 0 \,,
\end{eqnarray}
where $M_\chi (t) = g_S \, \phi(t) = m_\chi + g_S \Phi \cos ( m_\phi t
)$ .  
To solve Eq.~(\ref{eq:EOM_CHI}) we expand $\chi$ as
\begin{eqnarray}
    \chi(t, \vec x) = 
    \int \frac{d^3 k}{(2 \pi)^{3/2}} 
    \left[
        \chi_k (t) \, \hat a (\vec k)
        +
        \chi_k^\ast (t) \, \hat a^\dagger (-\vec k)
    \right] \, e^{i \vec k \cdot \vec x} \,,
\end{eqnarray}
where $\hat a (\vec k)$ and $\hat a^\dagger (\vec k)$ are annihilation
and creation operators, respectively, and they satisfy the commutation
relation $[ \hat a (\vec k_1) \,,~ \hat a^\dagger (\vec k_2) ] =
\delta^3 (\vec k_1 - \vec k_2)$.  The mode function $\chi_k$,
obeys the following equation of motion
\begin{eqnarray}
    \label{eq:EOM_CHIK}
    \ddot \chi_k(t) + \omega_k^2 (t) \, \chi_k(t) = 0 \,.
\end{eqnarray}
Here the time-dependent frequency is given by
\begin{eqnarray}
  \label{eq:OM_k}
    \omega_k^2 (t)
    =  k^2 + M_\chi^2 (t) 
    = \omega_\chi^2 + 2 \, g_S \, \Phi \, m_\chi  \cos (m_\phi t)
    + g_S^2 \, \Phi^2 \cos^2 (m_\phi t) \,,
\end{eqnarray}
where $\omega_\chi^2 = k^2 + m_\chi^2$ and $k=|\vec{k}|$.
In the method based on the Bogolyubov transformation, 
the growth of the mode function corresponds
to creations of $\chi$'s~\cite{Birrell}.  Indeed, the phase-space
distribution function of produced $\chi$'s is given by (see, {\it
  e.g.}, Ref.~\citen{Kofman:1994rk})
\begin{eqnarray}
    \label{eq:FCHI_0}
    f_\chi(t,k)
    =
    \frac{1}{2 \, \omega_k(t)}
    \Bigl[ \,
        |\dot{\chi}_k(t)|^2+\omega_k^2(t) \, |\chi_k(t)|^2 \,
    \Bigr]
    -\frac{1}{2} \,.
\end{eqnarray}
The number density of $\chi$ is then estimated as
\begin{eqnarray}
    n_\chi (t) = 
    \int \frac{d^3k}{(2 \pi)^3} \, f_\chi (t, k) \,.
\end{eqnarray}
As the initial conditions we take a plain wave solution such that
\begin{eqnarray}
    \label{eq:IC_CHIK}
    \chi_k (0) = \frac{1}{\sqrt{2 \, \omega_k (0)}} \,,~~~~
    \dot \chi_k (0) = - i \, \omega_k (0) \, \chi_k (0) \,.
\end{eqnarray}
In this case, $f_\chi(0,k)=0$ for all the momentum $k$
and the initial abundance of $\chi$ is zero.

Now we estimate the distribution function and 
the number density at the leading order of $g_S$.
For this purpose, we rewrite $\chi_k$
in the form (see, {\it e.g.}, Ref.~\citen{Kofman:1994rk})
\begin{eqnarray}
    \chi_k(t)
    = 
    \frac{\alpha_k(t)}{\sqrt{2 \, \omega_k(t)}} \,
    e^{-i \int_0^t dt_1 \, \omega_k(t_1) }
    +
    \frac{\beta_k(t)}{\sqrt{2 \, \omega_k(t)}} \,
    e^{+ i \int_0^t dt_1 \, \omega_k (t_1)} \,.
\end{eqnarray}
It is then found from Eq.~(\ref{eq:EOM_CHIK}) that
$\alpha_k$ and $\beta_k$ satisfy the equations 
\begin{eqnarray}
    \label{eq:A_S}
    \dot{\alpha}_k(t)
    \eqn{=} 
    \frac{\dot{\omega}_k(t)}{2 \, \omega_k(t)} \,
    e^{+ 2 i \int_0^t dt_1 \, \omega_k(t_1)} \,
    \beta_k(t) \,,
    \\
    \label{eq:B_S}
    \dot{\beta}_k(t)
    \eqn{=}
    \frac{\dot{\omega}_k(t)}{2 \, \omega_k(t)} \,
    e^{-2i\int_0^{t} dt_1 \, \omega_k (t_1)} \,
    \alpha_k(t) \,,
\end{eqnarray}
where $\alpha_k(0)=1$ and $\beta_k(0)=0$.
The coefficient functions obey the normalization condition
$|\alpha_k (t)|^2 - |\beta_k (t)|^2 = 1$.  In this case the
distribution function $f_\chi$ is written in terms of $\beta_k$ as
\begin{eqnarray}
    \label{eq:F_CHI}
   f_\chi (t, k) = |\beta_k(t)|^2 \,.
\end{eqnarray}

It should be noted that 
the factor $\dot \omega_k/(2 \, \omega_k)$ in
Eqs.~(\ref{eq:A_S}) and (\ref{eq:B_S}) is ${\cal O}(g_S)$:
\begin{eqnarray}
  \frac{\dot{\omega}_k(t)}{2\omega_k(t)}
  \eqn{=}
  \frac{{} - g_S \, \Phi \, m_\chi \, m_{\phi} \, \sin
  (m_{\phi}t) } { 2 \, \omega_\chi^2} + \mathcal{O}(g_S^2) \,.
\end{eqnarray}
The initial conditions of $\alpha_k$ and $\beta_k$ then show that the
leading contribution to $\beta_k$ is ${\cal O}(g_S)$, and
the time-dependence of $\alpha_k$ appears at ${\cal  O}(g_S^2)$.  
Therefore, $\beta_k$ at the leading order is obtained as
\begin{eqnarray}
    \label{eq:B_K}
    \beta_k(t) 
    \eqn{=}
    i \, \frac{g_S \, \Phi  \, m_\chi \, m_{\phi}}
    { 4 \, \omega_{\chi}^2 }
    \int_0^t dt_1 \,
    \bigl[ 
    e^{i ( m_\phi - 2 \, \omega_\chi ) t_1}
    -
    e^{- i ( m_\phi + 2 \, \omega_\chi ) t_1}
    \bigr]
    +\mathcal{O}(g_S^2) 
    \,.
\end{eqnarray}
This clearly shows that $\beta_k$, or equivalently $f_\chi$, 
causes oscillation for all the modes, 
except for the mode with $\omega_\chi = m_\phi/2$, \ie, with
the momentum $k = k_\ast$ where
\begin{eqnarray}
    k_\ast =  \frac{m_{\phi}}{2} \, \beta_\chi \,,
\end{eqnarray}
if $m_\chi < m_\phi/2$.%
\footnote{ For $m_\chi \gg m_\phi$, there is no growing mode.
  The study for such a case will be done in elsewhere.~\cite{AN}}
In this case the imaginary part of $\beta_k$ grows linearly with time
due to the cancellation of the phases between the $\phi$ oscillation
and the frequency of a pair of $\chi$.  It is also important to note
that $\omega_\chi=m_\phi/2$ for the growing mode suggests the energy
conservation in the process that $\phi$ at rest decays into a pair of
$\chi$ in the true vacuum.  For the growing mode the leading 
contribution to the occupation number is then estimated 
from Eq.~(\ref{eq:F_CHI}) as
\begin{eqnarray}
    \label{eq:FK_A}
    f_{\chi}(t, k_\ast)
    \simeq \left ( \frac{g_S \, \Phi \, m_\chi }{ m_{\phi} }\right)^2 \, t^2 
    \,,
\end{eqnarray}
where we have neglected the oscillation terms.
\begin{figure}[tb]
  \parbox[t]{\halftext}{%
    \includegraphics[scale=1.2]{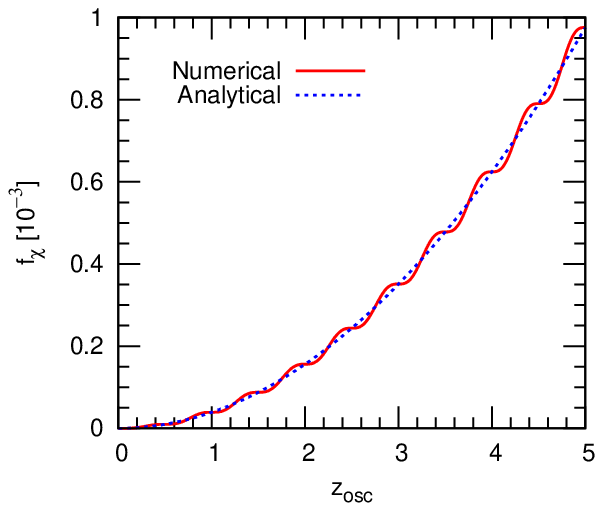}%
    \caption{ \label{fig:S_MODE_gm8}
      Evolution of the occupation number $f_\chi$ for the growing mode
      with $k = k_\ast$. The red solid line shows 
      the numerical result while
      the blue dashed line shows the analytical result~(\ref{eq:FK_A}).
      Here we take $g_S = 10^{-8}$ and $\beta_\chi = 0.5$.}
    }
  \hfill
  \parbox[t]{\halftext}{%
    \includegraphics[scale=1.2]{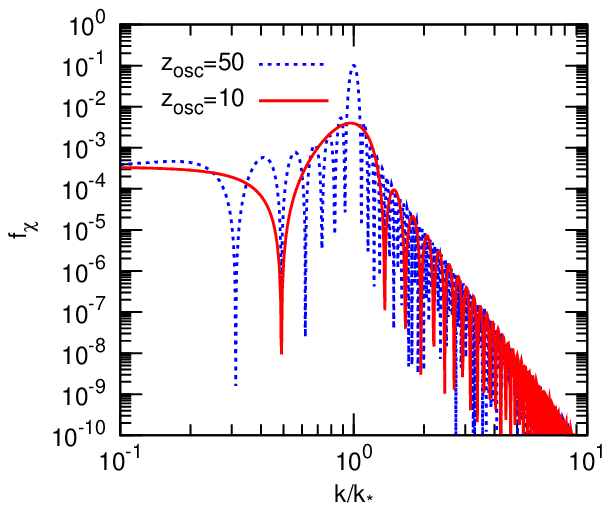}%
    \caption{ \label{fig:S_SPEC_gm8}
      Distribution function $f_\chi$ in momentum space
      $(k/k_\ast)$ at $z_{\rm osc} = 10$ (the red solid line)
      and at $z_{\rm osc} = 50$ (the blue dashed line).
      Here we take  $g_S=10^{-8}$ and $\beta_\chi=0.5$.}
  }
\end{figure}

In order to confirm the obtained results we numerically solve the
equation of motion (\ref{eq:EOM_CHIK}) and estimate the yield of
$\chi$.  As representative values we will take $m_\phi = 1.5 \times
10^{13}$ GeV and $\Phi = 3.4 \times 10^{19}$ GeV from now on.  In
Fig.~\ref{fig:S_MODE_gm8} we show the evolution of the occupation number
for the growing mode with $k=k_\ast$ in terms of the number of $\phi$
oscillation $z_{\rm osc} = m_\phi t/(2 \pi)$ by taking $g_S = 10^{-8}$
and $\beta_\chi = 0.5$ (\ie, $m_\chi \simeq 0.43 \, m_\phi$).  It is
seen that the analytical estimation in Eq.~(\ref{eq:FK_A})
successfully abstracts the characteristic behaviour of the occupation
number, $f_\chi (t, k_\ast) \propto t^2$.

The exact expression for the distribution function at ${\cal
  O}(g_S^2)$ is obtained from Eqs.~(\ref{eq:F_CHI}) and (\ref{eq:B_K}), 
which is given by
\begin{eqnarray}
    \label{eq:FCHI_g2}
    f_\chi (t,k) 
    \eqn{=}
    \frac{g_S^2 \, \Phi^2 \, m_\phi^2 \, m_\chi^2}{16 \, \omega_\chi^4} 
    I (t, \omega_\chi, m_\phi) \,,
\end{eqnarray}
where
\begin{eqnarray}
  \label{eq:ICHI_all}
    I (t, \omega_{\chi}, m_{\phi}) \eqn{=}
    \frac{2}{(m_\phi^2 - 4 \, \omega_\chi^2)^2}
    \Biggl\{
    3 m_\phi^2 + 4 \omega_\chi^2
    +( m_\phi^2 - 4 \omega_\chi^2) \cos( 2 m_\phi t)
    \nonumber \\
    \eqn{} \hspace{1ex}
    - 2 m_\phi 
    \Bigl[
        (m_\phi + 2 \omega_\chi)
        \cos \bigl( (m_\phi - 2 \omega_\chi)t \bigr)
        +
        (m_\phi - 2 \omega_\chi) 
        \cos \bigl( (m_\phi + 2 \omega_\chi)t \bigr)
    \Bigr]
    \Biggr\} \,.~~~~~~~
\end{eqnarray}
%
Fig.~\ref{fig:S_SPEC_gm8} shows the numerical results of the
distribution functions at $z_{\rm osc} = 10$ and $50$.  We have
confirmed that Eq.~(\ref{eq:FCHI_g2}) agrees with the
numerical results for the parameter choice in the figure.  It is seen
that $f_\chi$ has a peak at $k \simeq k_\ast$.  For the case of
$z_{\rm osc}= 10$ the peak is located at the momentum slightly smaller
than $k_\ast$ because of the effect of the oscillation terms, and such
an effect become negligible for larger $z_{\rm osc}$.  It is
interesting to note that the modes with $k > k_\ast$ are produced,
which are kinematically forbidden in the process where $\phi$ at rest
decays into a pair of $\chi$.  However, their occupation numbers are
highly suppressed and it scales as $k^{-8}$ or $k^{-6}$ when $2
z_{\rm osc}$ is integer or not, respectively.  On the other hand, for
the modes with $k \ll k_\ast$ the occupation number is independent on
$k$ and they oscillate around a constant value.  Furthermore, we can
also see from Fig.~\ref{fig:S_SPEC_gm8} that the typical width of the
peak in $f_\chi$ is inversely proportional to time.

\begin{figure}[t]
  \begin{center}
    \includegraphics[scale=1.2]{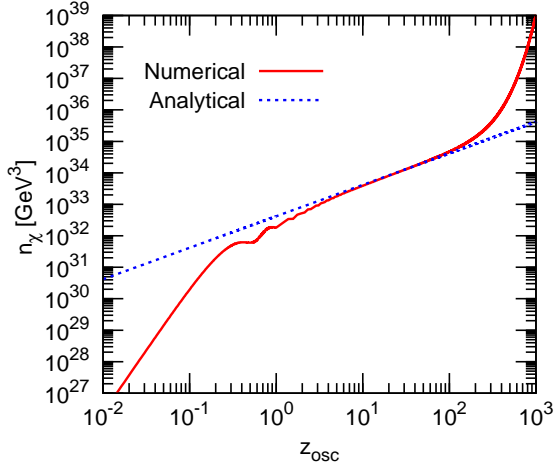}
  \end{center}
  \caption{ \label{fig:S_NUM_gm8}
    Evolution of the number density
    $n_\chi$ in terms of $z_{\rm osc}$.
    The red solid line shows the numerical result while
    the blue dashed line shows the analytical result~(\ref{eq:NCHI_ANA}).
    Here we take $g_S = 10^{-8}$ and $\beta_\chi = 0.5$.}
\end{figure}
The number density is then 
found at the leading order 
\begin{eqnarray}
    n_{\chi}(t)
    \simeq
    \frac{g_S^2 \, \Phi^2 \, m_\chi^2 \, \beta_\chi }{8\pi} \,  t \,.
    \label{eq:NCHI_ANA}
\end{eqnarray}
Here we have listed only the terms proportional to $t$ 
and neglected the oscillation terms.
The derivation of Eq.~(\ref{eq:NCHI_ANA}) is explained in Appendix~\ref{sec:AP1}. 
In Fig.~\ref{fig:S_NUM_gm8} we compare Eq.~(\ref{eq:NCHI_ANA}) with the
exact numerical result.
First, we observe that the number density grows as $t^4$ initially,
and there is a descrepancy between the leading ${\cal O}(g_S^2)$  
and numerical results.
After a few coherent oscillations, however, 
the number density is approaching to the ${\cal O}(g_S^2)$ result (\ref{eq:NCHI_ANA}),
and linearly grows in time with a small correction of oscillation.
Moreover, it should
be noted that the perturbative result (\ref{eq:NCHI_ANA}) coincides
with the naive result given in Eq.~(\ref{eq:Nchi_naive}), 
and hence $n_\chi$ at the leading order can be estimated by
the decays of non-relativistic $\phi$ into pairs of
$\chi$ after a few oscillation.  The same conclusion 
has been obtained in study of the narrow parametric resonance
by using the method of the density matrix.~\cite{Yoshimura:1996fk}
Finally, Fig.~\ref{fig:S_NUM_gm8} also shows that the perturbative result breaks down for $z_{\rm osc} \gtrsim 100$.
Therefore, we have clarified that the perturbative estimation
(as well as the naive estimation in Sec.~\ref{sec:Framework}) of the yield can be applied
only in the limited time interval.

\subsection{Production of fermion}
Next, we turn to consider the production of the Dirac fermion
$\psi$.  The equation of motion for $\psi$ is
\begin{eqnarray}
  \label{eq:EOM_PSI}
  \Bigl[ i \, \gamma^\mu \, \partial_\mu - M_\psi (t) \Bigr]
  \psi (t, \vec x) = 0 \,,
\end{eqnarray}
where $M_\psi(t) = g_F \, \phi (t) = m_\psi + g_F \, \Phi \, \cos
(m_\phi t )$.
We decompose $\psi$ as
\begin{eqnarray}
    \psi(t, \vec x) = \int \frac{d^3 k}{(2 \pi)^{3/2}}
    \sum_h 
    \left[ u_h (t, \vec k) \, \hat b_h (\vec k) 
        + v_h (t, \vec k) \, \hat d_h^\dagger (- \vec k)  
    \right] \, e^{i \vec k \cdot \vec x} \,,
\end{eqnarray}
where the summation is taken over helicity $h=\pm$, and $v_h (t, \vec
k) = u_h^c (t, - \vec k)$.  $\hat b_h$ and $\hat d_h$ are annihilation
operators of particle and anti-particle, respectively, and they
satisfy the anti-commutation relations $\{ \hat b_{h_1}( \vec
k_1),\hat b_{h_2}^\dagger ( \vec k_2) \} = \{ \hat d_{h_1}( \vec
k_1),\hat d_{h_2}^\dagger ( \vec k_2) \} = \delta_{h_1 h_2}\delta^3
(\vec k_1 - \vec k_2)$.  We write the wave function $u_h$ as
\begin{eqnarray}
    u_h (t, \vec k) = 
    \left(
        \begin{array}{c}
            P_{h}(t, k) \\
            Q_{h}(t, k)
        \end{array}
    \right) \otimes
    \varphi_h (\vec k) \,,
\end{eqnarray}
where the helicity eigenfunction satisfies
$\vec k \, \varphi_h (\vec k) = h k \, \varphi_h
(\vec k)$.  Note that the mode functions $P_h$ and $Q_h$ obey
the normalization condition $| P_h (t, k) |^2 + | Q_h (t, k) |^2 = 1 $.
It is then found from Eq.~(\ref{eq:EOM_PSI}) that 
the equation of motion $P_h$ is 
\begin{eqnarray}
    \label{eq:EOM_PH}
    \ddot P_h (t, k) + \tilde \omega_k^2 (t)\, P_h (t, k) = 0 \,,
\end{eqnarray}
where $\tilde \omega_k$ is given by
\begin{eqnarray}
    \tilde \omega_k^2 (t) 
    =
    \omega_k^2 (t) + i \, \dot {M}_\psi (t) 
    =
    k^2 + M_\psi^2(t) + i \, \dot {M}_\psi (t) \,.
\end{eqnarray}
As in the case of the scalar production, we assume a plain wave solution
initially and impose the conditions
\begin{eqnarray}
    \label{eq:IC_PH}
    P_h (0, k) = 
    \sqrt{ 
      \frac{\omega_k (0) + M_\psi (0)}{ 2 \, \omega_k(0) } 
    }\,,~~~~
    \dot P_h (0,k) =
    - i \, \omega_k (0) \, P_h (0,k) 
    \,.
\end{eqnarray}
It is then seen that the equations of motion and the initial
conditions for the $h = \pm$ states are the same, and so $P_+ (t,k) =
P_-(t,k)$.  On the other hand, $Q_h$ is obtained from $P_h$ as
\begin{eqnarray}
    Q_h (t,k) = - i \, \frac{h}{k}
    \left[ \dot P_h (t,k) + i M_\psi (t) \, P_h (t, k) \right] \,. 
\end{eqnarray}
This means that $Q_+ (t,k) = - \, Q_-(t,k)$.
The distribution function of $\psi$ can be written in terms
of the mode functions as 
(see, {\it e.g.}, Ref.~\citen{Garbrecht:2002pd})
\begin{eqnarray}
  \label{eq:FPSI}
    f_\psi(t, k) = 
    \frac 12 - \frac{\Omega_h (t,k)}{2 \, \omega_k (t) } \,,
\end{eqnarray}
where $\Omega_h$ is defined by
\begin{eqnarray}
  \label{eq:BOM}
    \Omega_h (t, k) \eqn{=}
    - h k 
    \Bigl[ P_h (t, k) Q_h^\ast (t,k) + Q_h(t,k) P_h^\ast (t,k) \Bigr]
    + M_\psi (t) \Bigl[ 
    \bigl| P_h(t,k) \bigr|^2 - \bigl| Q_h(t,k) \bigr|^2 \Bigr] 
    \nonumber \\
    \eqn{=}
    - 2 \, \mbox{Im} 
    \Bigl[
    P_h^\ast (t,k) \, \dot P_h(t,k)
    \Bigr] - M_\psi (t)
    \,.
\end{eqnarray}
This clearly shows that the distribution functions for two helicity
states are exactly the same.  Finally, the number density of $\psi$ is
given by
\begin{eqnarray}
  n_\psi (t) =  4 \int \frac{d^3k}{(2 \pi)^3} \,
  f_\psi (t, k) \,.
\end{eqnarray}
Here a factor of four counts the number of internal degrees
of freedom of $\psi$.

We then turn to estimate the leading contribution
to the yield of $\psi$ when $g_F$ is very small.  
Since $P_+=P_-$,
the index $h$ will be implicit from now on.  In order to solve
Eq.~(\ref{eq:EOM_PH}) we express $P$ as
\begin{eqnarray}
    P (t,k) = 
    \frac{A_{k}(t)}{\sqrt{2 \, \tilde \omega_k(t)}} \,
    e^{ - i \int_0^t dt_1 \, \tilde \omega_k (t_1) }
    +
    \frac{B_{k}(t)}{\sqrt{2 \, \tilde \omega_k(t)}} \,
    e^{ + i \int_0^t dt_1 \, \tilde \omega_k (t_1) }\,.
\end{eqnarray}
In this case the coefficients $A_k$ and $B_k$ obey the coupled
equations
\begin{eqnarray}
    \label{eq:EOM_AF}
    \dot A_k(t) \eqn{=} 
    \frac{ \dot{ \tilde \omega}_k (t)}{ 2 \, \tilde \omega_k (t)}
    \exp \left[ + 2 i \int_0^t dt_1 \, \tilde \omega_k (t_1) \right]
    B_k(t) \,,
    \\
    \label{eq:EOM_BF}
    \dot B_k(t) \eqn{=} 
    \frac{ \dot{ \tilde \omega}_k (t)}{ 2 \, \tilde \omega_k (t)}
    \exp \left[ - 2 i \int_0^t dt_1 \, \tilde \omega_k (t_1) \right]
    A_k (t) \,,
\end{eqnarray}
and their initial values are found from Eq.~(\ref{eq:IC_PH}) as
\begin{eqnarray}
    A_k (0) \eqn{=}
    \Bigl[ m_\psi + g_F \Phi + 
    \bigl(
        \omega_\psi^2 + 2 g_F \Phi m_\psi + g_F^2 \Phi^2 \bigr)^{1/2}
    \Bigr]^{1/2} \,,
    ~~~
    B_k (0) = 0 \,,
    \label{eq:IC_F}
\end{eqnarray}
where $\omega_\psi^2 = k^2 + m_\psi^2$ is the frequency of $\psi$ in
the true vacuum $\Phi = 0$.  From now on let us find solutions of
$A_k$ and $B_k$ in power series of coupling $g_F$ as we did in the
$\chi$ production.

The leading term of $B_k$ is found to be ${\cal O}(g_F)$, which can be
obtained as
\begin{eqnarray}
  \label{eq:BK_G1}
    B_k (t) 
    \eqn{=} 
    \int_0^t dt_1 \, 
    \frac{ \dot{ \tilde \omega}_k (t_1)}{ 2 \, \tilde \omega_k (t_1)} \,
    e^{ - 2 i \int_0^{t_1} dt_2 \, \tilde \omega_k (t_2) } \,
    A_k (t_1) 
    =
    - i \, \frac{g_F \, \Phi \, m_\phi}{8 \, \omega_\psi^2} \, 
    A_k^{(0)} \, I^\psi_1 (t,k)
    + {\cal O}(g_F^2) \,,
\end{eqnarray}
where $A_k^{(0)} =\sqrt{ \omega_\psi + m_\psi }$ denotes the ${\cal O}(g_F^0)$
term of $A_k$ which is independent on time, and we have introduced
\begin{eqnarray}
    \label{eq:I_PSI1}
    I^\psi_1(t,k) =
    \int_0^t dt_1 
    \Bigr[
        (m_\phi - 2 m_\psi) \, e^{i (m_\phi - 2 \omega_\psi) t_1}
        +
        (m_\phi + 2 m_\psi) \, e^{-i (m_\phi + 2 \omega_\psi) t_1}
    \Bigl] \,.
\end{eqnarray}
Then, we can again see that $B_k$'s oscillate with time for all the
modes expect for the mode with $\omega_\psi = m_\phi/2$, \ie, with the
momentum
\begin{eqnarray}
    k_\ast = \frac{m_\phi}{2} \beta_\psi \,,
\end{eqnarray}
if $m_\psi < m_\phi/2$. Notice that it corresponds to
the momentum of $\psi$ in the decay process $\phi \to \psi + \overline
\psi$.  For this growing mode we find that
\begin{eqnarray}
    \label{eq:BK_GROW}
    B_{k_\ast}(t) = - i \, \frac{g_F \, \Phi \, A_{k_\ast}^{(0)}}
    {2 \, m_\phi} \,
    (m_\phi - 2 \, m_\psi) \, t + \cdots \,,
\end{eqnarray}
and increases linearly in time (with corrections of
oscillations), which is consequence of the cancellation of phases
between the $\phi$ oscillation and the frequency of a pair of
$\psi$.  Moreover,  $A_{k_\ast}$ increases as $t^2$ at ${\cal O}(g_F^2)$,
which can be seen as follows:  We find from Eq.~(\ref{eq:EOM_AF})
that 
\begin{eqnarray}
  \label{eq:AK_G2}
    A_{k}(t) 
    \eqn{=} 
    A_k(0) 
    + 
    \int_0^t dt_1 \, 
    \frac{ \dot{ \tilde \omega}_k (t_1)}{ 2 \, \tilde \omega_k (t_1)} \,
    e^{ + 2 i \int_0^{t_1} dt_2 \, \tilde \omega_k (t_2) } \,
    B_k (t_1)  
    \nonumber \\
    \eqn{=}
    A_k(0)
    - i \, \frac{ g_F \, \Phi \, m_\phi}{8 \, \omega_\psi^2}  \,
    I^\psi_2 (t, k) 
    + \cdots \,,
\end{eqnarray}
where 
\begin{eqnarray}
    \label{eq:I_PSI2}
    I^\psi_2 (t, k)
    =
    \int_0^t dt_1 \,
    \Bigl[
    ( m_\phi - 2 m_\psi ) \, e^{i (m_\phi + 2 \omega_\psi) t_1 }
    +
    ( m_\phi + 2 m_\psi ) \, e^{- i (m_\phi - 2 \omega_\psi) t_1 }
    \Bigr] \, B_k(t_1) \,.
\end{eqnarray}
By using Eq.~(\ref{eq:BK_GROW}) $A_{k_\ast}$ is given by
\begin{eqnarray}
    A_{k_\ast}(t) = A_{k_\ast} (0)
    - \frac{g_F^2 \, \Phi^2 \, \beta_\psi^2 \, A_{k_\ast}^{(0)} }{8} \, 
    t^2 + \cdots \,.
\end{eqnarray}
Therefore, the leading  ${\cal O}(g_F^2)$ contribution
to the occupation number of the mode $k=k_\ast$ is
found from Eqs.~(\ref{eq:FPSI}) and (\ref{eq:BOM})
\begin{eqnarray}
    \label{eq:FPSI_ANA}
    f_\psi (t, k_\ast )
    \simeq \left( \frac{g_F \, \Phi \, \beta_\psi}{2} \right)^2 \, t^2 \,.
\end{eqnarray}
by neglecting the oscillation terms.  

\begin{figure}[t]
  \parbox[t]{\halftext}{%
    \includegraphics[scale=1.2]{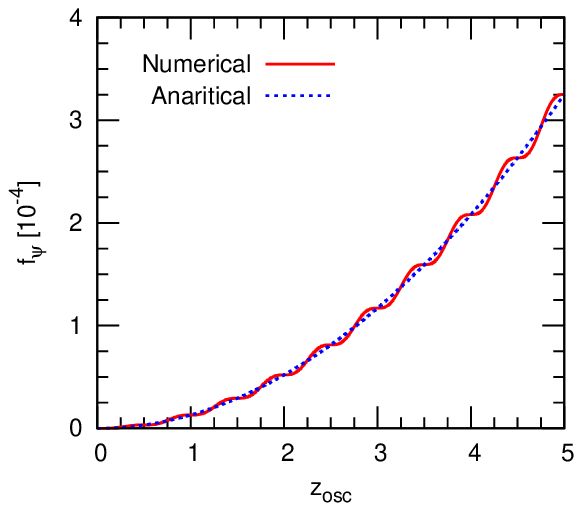}%
    \caption{ \label{fig:F_MODE_gm8}
      Evolution of the occupation number $f_\psi$ for the growing mode
      with $k = k_\ast$. The red solid line shows 
      the numerical result while
      the blue dashed line shows the analytical result~(\ref{eq:FPSI_ANA}).
      Here we take $g_F = 10^{-8}$ and $\beta_\chi = 0.5$.}
    }
  \hfill
  \parbox[t]{\halftext}{%
    \includegraphics[scale=1.2]{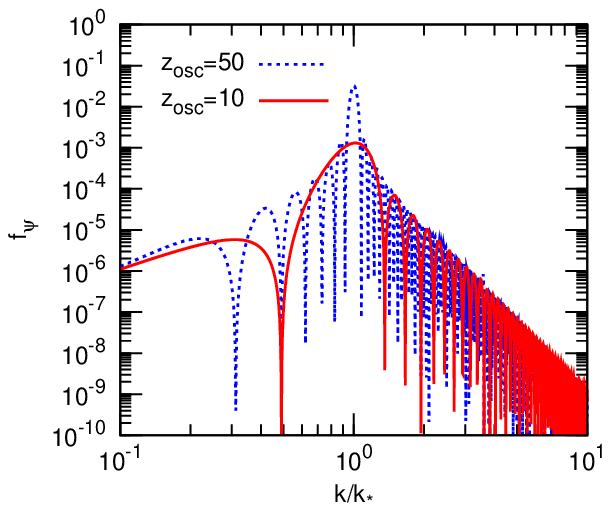}%
    \caption{ \label{fig:F_SPEC_gm8}
      Distribution function $f_\psi$ in momentum space
      $(k/k_\ast)$ at $z_{\rm osc} = 10$ (the red solid line)
      and at $z_{\rm osc} = 50$ (the blue dashed line).
      Here we take  $g_F=10^{-8}$ and $\beta_\chi=0.5$.}
  }
\end{figure}
We then compare the above result with the numerical solution of
Eq.~(\ref{eq:EOM_PH}) which includes the higher order terms of $g_F$
and the oscillation terms.  We show in Fig.~\ref{fig:F_MODE_gm8} the
evolution of the occupation number for the growing mode by taking
$g_F=10^{-8}$ and $\beta_\psi =0.5$ (\ie, $m_\psi \simeq 0.43 \, m_\phi$).
We can see that the perturbative result~(\ref{eq:FPSI_ANA}) gives a
good approximation for $f_\psi$, and it increases at $t^2$ with
corrections from the oscillation terms.

The exact expression for the distribution function at the leading
$g_F^2$ order is found from Eqs.~(\ref{eq:BK_G1}) and (\ref{eq:AK_G2})
as
\begin{eqnarray}
    \label{eq:FPSI_g2}
    f_\psi (t,k) 
    \eqn{=}
    \frac{g_F^2 \, \Phi^2 \, m_\phi^2 \, k^2}{16\, \omega_\psi^4} 
    I (t,\omega_\psi, m_\phi) \,,
\end{eqnarray}
where $I (t,\omega_\psi, m_\psi)$ is given by Eq.~(\ref{eq:ICHI_all}).
Fig.~\ref{fig:F_SPEC_gm8} shows the numerical estimation of the
distribution function $f_\psi$.  Notice that we have confirmed that
the analytical result in Eq.~(\ref{eq:FPSI_g2}) agrees with
the numerical one, as in the case of the scalar production.  
It is seen that $f_\psi$ has a peak at $k \simeq k_\ast$
and its height scales as $t^2$ as expected.  Further, the figure shows
that the typical width of the peak becomes narrow as $\Delta
k (t) \propto 1/t$.  It is also interesting to compare
Eq.~(\ref{eq:FCHI_g2}) with Eq.~(\ref{eq:FPSI_g2}).  The differences
between $f_\psi$ and $f_\chi$ are in the prefactor ($k
\leftrightarrow m_\chi$) and in the arguments of the function $I$.
Because of these differences, $f_\psi$ for the modes 
$k \ll k_\ast$ does dependent on $k$ 
in contrast to scalar production 
and the suppression of $f_\psi$ for $k \gg k_\ast$ is relaxed.
Moreover, the former difference provides the
additional $\beta_\psi^2$ factor for the number density of produced $\psi$.   
\begin{figure}[t]
    \begin{center}
    \includegraphics[scale=1.2]{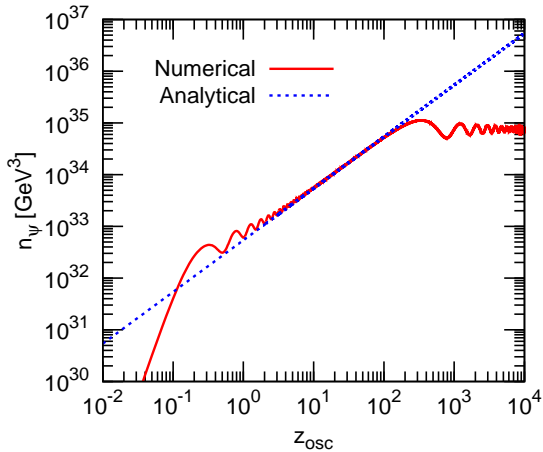}
    \end{center}
    \caption{ Evolution of the number density
      $n_\psi$ in terms of $z_{\rm osc}$.  The red solid line
      shows the numerical result while the blue dashed line shows the
      analytical result~(\ref{eq:NPSI_ANA}).  Here we take
      $g_F = 10^{-8}$ and $\beta_\psi = 0.5$.  }
    \label{fig:F_NUM_gm8} 
\end{figure}
As shown in Appendix~\ref{sec:AP1},
the leading contribution to
the number density of $\psi$ is given by
\begin{eqnarray}
    \label{eq:NPSI_ANA}
    n_\psi(t) 
    \simeq \frac{g_F^2 \, \beta_\psi^3 \, \Phi^2 \, m_\phi^2}{8 \pi} \, t \,,
\end{eqnarray}
by neglecting the oscillation terms. 
It is then found that the number density becomes independent
on $m_\psi$ for $m_\psi \ll m_\phi$
(as long as $\Phi \ll \vev{\phi}$).
Notice that it coincides with
the naive result in Eq.~(\ref{eq:Npsi_naive}).  

In Fig.~\ref{fig:F_NUM_gm8} the evolution of the number density
$n_\psi$ is also shown.  We can see that the numerical result is
approaching to the analytical estimate~(\ref{eq:NPSI_ANA}) after a few
oscillations of $\phi$.  Therefore, the leading ${\cal O}(g_F^2)$
contribution to the $\psi$ yield is also described by the decays of
non-relativistic $\phi$ particles into pairs of $\psi$ and
$\overline{\psi}$.
However, it is also found from Fig.~\ref{fig:F_NUM_gm8} that the  $\mathcal{O}(g_F^2)$ result
in Eq.~(\ref{eq:NPSI_ANA}) breaks down in sufficiently later times, 
say $z_{\rm osc} \gtrsim 100$, as in the scalar production.
This issue will be discussed in the next section.
\section{Non-perturbative Corrections}
\label{sec:Time_Ave}
We have so far estimated the leading contributions to the yields of
$\chi$ and $\psi$ from the coherent oscillation.  
When the amplitude of the oscillation is small
enough, the yields are induced at the order of ${\cal
  O}(g_{S,F}^2)$.  In this case there exists the growing mode with
$k=k_\ast$ if $m_{\chi, \psi} < m_\phi/2$ and its occupation number
grows at the rate $t^2$ after a small number of oscillations.  The number
density $n_{\chi, \psi}$ becomes proportional to $t$, which is consistent
with the naive estimation in Eqs.~(\ref{eq:Nchi_naive}) and
(\ref{eq:Npsi_naive}) based on the decays of non-relativistic
particles.  It is, however, found that the perturbative result
eventually fails to describe the evolution of the yield at later
times.

\begin{figure}[t]
  \parbox[t]{\halftext}{%
    \includegraphics[scale=1.2]{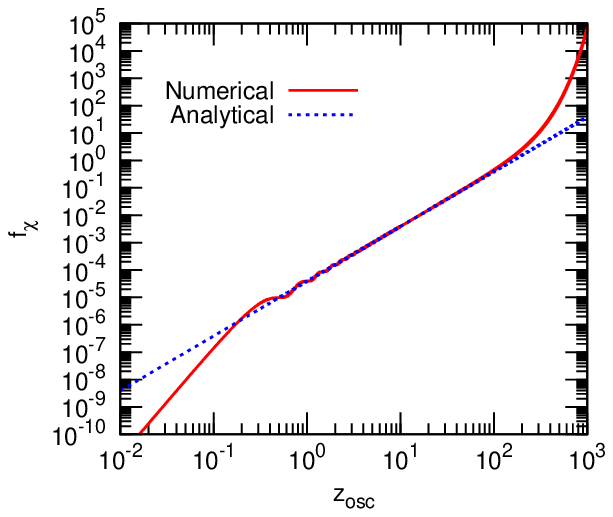}%
    \caption{ \label{fig:S_MODE_gm8_L}
      Same as Fig.~\ref{fig:S_MODE_gm8} except for the 
      range of $z_{\rm osc}$.}
  }
  \hfill
  \parbox[t]{\halftext}{%
    \includegraphics[scale=1.2]{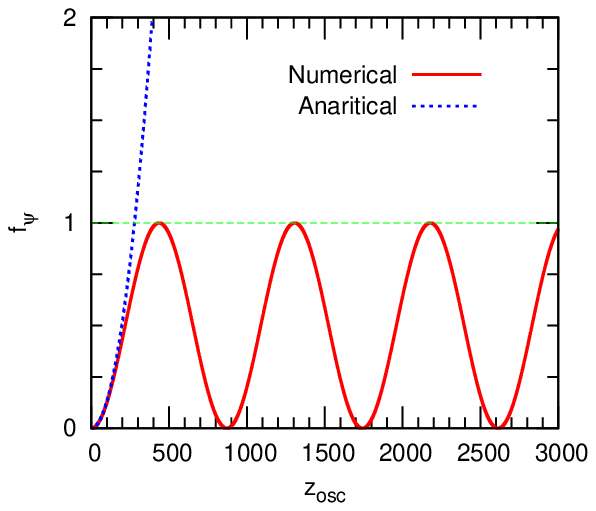}
    \caption{ \label{fig:F_MODE_gm8_L}
      Same as Fig.~\ref{fig:F_MODE_gm8} except for 
      the range of $z_{\rm osc}$.}
  }
\end{figure}
As for the $\chi$ production, Fig.~\ref{fig:S_MODE_gm8_L} shows how the
occupation number for the growing mode evolves
at later epoch.  We can see that $f_\chi(t, k_\ast)$ starts to grow
exponentially, which is different from the ${\cal O}(g^2_S)$ result
given in (\ref{eq:FK_A}).  As already pointed out in
Ref.~\citen{Kofman:1994rk}, this discrepancy comes from the
non-perturbative effect due to the narrow resonance at the preheating
stage.  Such an explosive production is reflected the statistical
property of $\chi$, \ie, the effect of the Bose condensation.
Then, the number density of $\chi$ also grows exponentially 
as also shown in Fig.~\ref{fig:S_NUM_gm8}.
On the other hand, with regard to the $\psi$ production, the later
evolution of $f_\psi (t, k_\ast)$ is shown in
Fig.~\ref{fig:F_MODE_gm8_L}.  In this case the ${\cal O}(g^2_{F})$
result in (\ref{eq:FPSI_ANA}) becomes inconsistent in the end
and it oscillates between zero and unity.  This
oscillation behaviour had already been observed in
Ref.~\citen{Greene:1998nh}, which is a consequence of 
the Pauli-blocking effect, \ie, $f_\psi$ is forbidden to exceed unity.  
Accordingly, the number density stops to grow at some point as also 
shown in Fig.~\ref{fig:F_NUM_gm8}.

The importance of these non-perturbative corrections have already 
been addressed in various cases of the preheating process.
It should be stressed that such corrections become significant 
eventually even if the coupling constant $g_{S,F}$ is extremely small.  
In general, it is a difficult task
to derive the analytical expression for the yield including
the non-perturtative effect.
The previous works~\cite{Kofman:1994rk,Yoshimura:1995gc,Yoshimura:1996fk} 
had investigated  by using the knowledge of the Mathieu function,~\cite{Lauchlan:1961}
since the solution of (\ref{eq:EOM_CHIK}) is approximately given by this function.
Here we utilize another method, 
the time averaging method, which is familiar in the
study of the non-linear differential equations.~\cite{NAYFEH}
From now on we will demonstrate that the evolution of the occupation 
number for the growing mode can be successfully described by
this method.
Note that the time averaging method has
already been used to describe the resonance structure of Mathieu
function~\cite{Traschen:1990sw}, where the authors 
have focused only on the scalar production, and have obtained
the Mathieu characteristic exponent of the first instability band  
and Eq.~(\ref{eq:Fchi_TimeAve}) shown in below.

We will demonstrate below a simpler application of the time averaging method
 together with the method of variation of parameters.
This allows us to clearly derive the analytical forms of 
the mode function and the occupation number for the growing mode.
Furthermore, our approach is applicable not only to the scalar production
but also to the fermion production as shown in below.

\subsection{Growing mode of scalar}
Let us recall the equation of motion for $\chi_k$ (\ref{eq:EOM_CHIK})
 for the growing mode $k=k_\ast$
\begin{eqnarray}
  \label{eq:chi_k}
  \frac{d^2 \chi_{k_\ast}}{d \tau^2} 
  +
  \left[
    1 + 2 \, q_\chi \, \sqrt{1- \beta_\chi^2 } \, \cos (2 \tau)
    + q_\chi^2 \, \cos^2 ( 2 \tau ) 
  \right] \, \chi_{k_\ast} = 0 \,,
\end{eqnarray}
where we have introduced $\tau= m_\phi t /2$, and
$q_\chi = 2 g_S \Phi/m_\phi$.  We shall solve this equation by using
the methods of variation of parameters and time averaging.  For this
purpose, we introduce $u_1$ and $u_2$ by
\begin{eqnarray}
  \chi_{k_\ast}(\tau) = y_1(\tau) \, u_1 (\tau) 
  + y_2 (\tau) \, u_2(\tau) \,,
\end{eqnarray}
with the condition
\begin{eqnarray}
  \label{eq:C_VP}
  y_1(\tau) \, \frac{d u_1(\tau)}{d \tau}
  + y_2 (\tau) \, \frac{ d u_2 (\tau)}{d \tau} = 0 \,,
\end{eqnarray}
where $y_1 = \cos \tau$ and $y_2 = \sin \tau$ are
the solutions for Eq.~(\ref{eq:chi_k}) with $q_\chi =
0$.  We then obtain the equations for $u_1$ and $u_2$ as
\begin{eqnarray}
  \frac{d}{d\tau}
  \left(
    \begin{array}{c}
      u_1 \\ u_2
    \end{array}
  \right)
  \eqn{=}
  \left[ 2 \, q_\chi \, \sqrt{1-\beta_\chi^2}\, \cos(2 \tau)
  + q^2_\chi \, \cos^2 (2 \tau) \right]
  \nonumber \\
  \eqn{}\times
  \left(
    \begin{array}{c c}
      \sin \tau \, \cos \tau & \sin^2 \tau \\
      - \cos^2 \tau & - \sin \tau \, \cos \tau 
    \end{array}
  \right)
  \left(
    \begin{array}{c}
      u_1 \\ u_2
    \end{array}
  \right) \,.
\end{eqnarray}
Now we are interested in the characteristic behaviour of
$\chi_{k_\ast}$ due to the non-perturbative effect and, as shown
below, its typical time scale is given by $\tau \sim 1/q_\chi$ which
is much longer than the period of the $\phi$ oscillation in the weak
coupling limit (say, $q_\chi \ll 1$).  In this situation we can apply
the time averaging method which extract the underlying behaviour over a long time scale
by integrating out the effects of the rapid $\phi$ 
oscillations.
The $u_1$ and $u_2$ averaged over the
oscillation period are denoted by $\overline u_1$ and $\overline u_2$,
and they satisfy
\begin{eqnarray}
  \label{eq:TAV_CHI}
  \frac{d}{d\tau}
  \left(
    \begin{array}{c}
      \overline u_1 \\
      \overline u_2
    \end{array}
  \right)
  = 
  - \frac{q_\chi \sqrt{1- \beta_\chi^2}}{2}
  \left(
    \begin{array}{c c}
      0 & 1 \\
      1 & 0
    \end{array}
  \right)
  \left(
    \begin{array}{c}
      \overline u_1 \\ \overline u_2
    \end{array}
  \right)\,.
\end{eqnarray}
Here and hereafter, we neglect the contributions of higher order of
$q_\chi$.  By using the initial
conditions (\ref{eq:IC_CHIK}) at ${\cal O}(g_S^0)$,
$\chi_{k_\ast}$ after the time averaging 
is obtained as
\begin{eqnarray}
  \label{eq:TAV_CHIK}
  \chi_{k_\ast}(\tau) 
  \eqn{=}
  y_1 (\tau ) \overline u_1 (\tau) +
  y_2 (\tau ) \overline u_2 (\tau) 
  \nonumber \\
  \eqn{=}
  \frac{1}{\sqrt{m_\phi}}
  \left[ 
    e^{- i \tau} \, 
    \cosh \left( \frac{q_\chi \sqrt{1-\beta_\chi^2}}{2} \tau \right)
    +
    i \,
    e^{+ i \tau} \, 
    \sinh \left( \frac{q_\chi \sqrt{1-\beta_\chi^2}}{2} \tau \right)
  \right] \,.
\end{eqnarray}
Finally, Eq.(\ref{eq:FCHI_0}) gives the occupation number
for the growing mode with $k=k_\ast$ as
\begin{eqnarray}
  \label{eq:Fchi_TimeAve}
  f_\chi (t, k_\ast) 
  =
  \sinh^2 \left( 
    \frac{g_S \Phi}{2} \, \sqrt{1- \beta_\chi^2} \, t
  \right)
  =
  \sinh^2 \left( 
    \frac{g_S \, \Phi \, m_\chi}{m_\phi} \, t
  \right)
  \,.
\end{eqnarray}

We have checked that the obtained result can describe the exact
numerical one in Fig.~\ref{fig:S_MODE_gm8_L} apart from tiny
corrections of oscillation.  Interestingly, this correctly reproduces
the initial behaviour in Eq.~(\ref{eq:FK_A}) for $ 1/m_\phi \ll t \ll
t_{\rm cr} = m_\phi/(g_S \Phi m_\chi)$.  When $g_S=10^{-8}$ and
$m_\chi \simeq 0.43 m_\phi$ shown in Fig.~\ref{fig:S_MODE_gm8_L}, the
critical time is estimated to be $z_{\rm cr}\simeq 160$ 
(\ie, $t_{\rm cr}\simeq 7 \times 10^{-11}\GeV^{-1}$).
Note that the exact numerical
estimation of $f_\chi$ shows $f_\chi \propto t^4$ for the beginnings
of production (within one oscillation), which can not be explained by
this result.  On the other hand, for $t \gg t_{\rm cr}$, the
occupation number grows exponentially as $\exp ( 2 g_S \Phi m_\chi t
/m_\phi)$.  This exponent is consistent with the result from
the narrow parametric resonance.~\cite{Kofman:1994rk}
Notice again that such non-perturbative correction becomes
significant for $t \gtrsim t_{\rm cr}$ even if the coupling $g_S$ is
extremely small.  Accordingly, the number density starts to grow
exponentially at $t \simeq t_{\rm cr}$.

It is important to note that the mode function for the growing mode in
Eq.~(\ref{eq:TAV_CHIK}) satisfies the exact scaling
property~\cite{Mostepanenko:1974} in its time evolution, which is a
consequence of the periodicity of the $\phi$ oscillation.
To see this point, let us first recall the exact equation of motion for
$\chi_k$ in Eq.~(\ref{eq:EOM_CHIK}) and denote by $\chi_{k}^{(1)}(t)$ and
$\chi_{k}^{(2)}(t)$ its two linearly independent solutions with
the initial conditions 
$\chi_{k}^{(1)}(0)=\dot{\chi}_{k}^{(2)}(0)=1$ and 
$\dot{\chi}_{k}^{(1)}(0)=\chi_{k}^{(2)}(0)=0$.
In this case, $\chi_k$ can be written without loss of generality as
\begin{equation}
  \label{eq:CHIK_SCALE}
  \chi_k(t)
  =
  \chi_k(0)
  \left[ 
    \chi_k^{(1)}(t) - i \omega_k(0) \chi_k^{(2)}(t)
  \right] \,.
\end{equation} 
Due to the periodicity of $\omega_k(t)$, 
the independent solutions satisfy the 
exact scaling property~\cite{Mostepanenko:1974}
\begin{eqnarray}
  \label{eq:MAP_CHIK}
  \left(
    \begin{array}{c}
      \chi_k^{(1)}(t+T) \\
      \chi_k^{(2)}(t+T)
    \end{array}
  \right)=
  \left(
    \begin{array}{c c}
      \chi_k^{(1)}(T) & \dot{\chi}_k^{(1)}(T) \\
      \chi_k^{(2)}(T) & \dot{\chi}_k^{(2)}(T)
    \end{array}
  \right)
  \left(
    \begin{array}{c}
      \chi_k^{(1)}(t) \\
      \chi_k^{(2)}(t)
    \end{array}
  \right) \,,
\end{eqnarray}
where $T$ is an oscillation period of $\phi(t)$ 
($\ie, \, \omega_k(t+T)=\omega_k(t)$).
This shows that we can extrapolate the mode function at $t=n T$ by using 
the solution at $t=T$ recursively.
From this exact property, 
the occupation number of $\chi$ at the time $t=nT$ 
is written by 
\begin{eqnarray}
    f_\chi(nT ,k)
  \eqn{=}
  \frac{1}{4\omega_k^2(T)}
  \left(\frac{\sinh  \left(n D\right)}{\sinh \left(D\right)}\right)^2
  \left(\dot{\chi}_k^{(1)}(T)+\omega_k^2(T)\chi_k^{(2)}(T)\right)^2 \nonumber\\
  \eqn{=}\left[ \frac{\sinh  \left(n D\right)}{\sinh  \left(\left(n-1\right) D\right)} \right]^2 f_\chi \bigl( (n-1) T , k \bigr) \,,
\label{eq:FCHI_NT}
\end{eqnarray}
where $D\equiv\cosh^{-1}(\chi_k^{(1)}(T))$.
Therefore, the occupation number at $t=n T$ 
can be obtained by 
using the mode function at $t=T$.

Now, we examine whether the occupation number for the 
growing mode by the time averaging method~(\ref{eq:TAV_CHIK})
satisfy these scaling properties or not.
It should be noted that our result can be 
written as Eq.(\ref{eq:CHIK_SCALE}) with
\begin{eqnarray}
  \label{eq:DEC_CHIK1}
  \chi_{k_\ast}^{(1)}(t) 
  \eqn{=} 
  \cosh \left(\frac{g_S \, \Phi \, m_\chi}{m_\phi}t\right) 
  \cos\left(\frac{m_\phi t}{2}\right)
  -
  \sinh \left(\frac{g_S\, \Phi \, m_\chi}{m_\phi}t\right) 
  \sin\left(\frac{m_\phi t}{2}\right)\, ,
  \\
  \label{eq:DEC_CHIK2}
  \chi_{k_\ast}^{(2)}(t) 
  \eqn{=} 
  \frac{2}{m_\phi}
  \left[ 
    \cosh \left(\frac{g_S \, \Phi \, m_\chi}{m_\phi}t\right)
    \sin\left(\frac{m_\phi t}{2}\right)
    -
    \sinh \left(\frac{g_S \, \Phi \, m_\chi}{m_\phi}t\right) 
    \cos\left(\frac{m_\phi t}{2}\right)
  \right] \, .
\end{eqnarray}
Here we have taken the mode function at the initial time 
as a free field.  
We can show that these functions satisfy
the exact scaling property~(\ref{eq:MAP_CHIK}).
See the details in Ref.~\citen{AN}.
Moreover, substituting Eqs.~(\ref{eq:DEC_CHIK1})
and (\ref{eq:DEC_CHIK2}) into Eq.~(\ref{eq:FCHI_NT}), we obtain the
occupation number at $t=n T$ with exact scaling property as
\begin{equation}
  \label{eq:FCHI_NT_SCL}
  f_\chi(nT , k_\ast) 
  = 
  \sinh^2\left( \frac{g_S\, \Phi \, m_\chi}{m_\phi} \, n T\right)\,.
\end{equation}
This result is consistent with Eq.~(\ref{eq:Fchi_TimeAve}).
Therefore, the analytical result for the occupation number 
obtained by the time averaging method
can satisfy the exact relation 
in its evolution coming from the periodicity of 
the $\phi$ oscillation.

\subsection{Growing mode of fermion}
Next, we turn to consider the occupation number
of $\psi$ with $k = k_\ast$ at later time.
The equation of motion for the mode function is now given by
\begin{eqnarray}
  \label{eq:pk}
  \frac{d^2 P(\tau)}{d\tau^2}
  +
  \left[
    1 + 2 \, q_\psi \, \sqrt{1- \beta_\psi^2} \, \cos (2 \tau )
    - 
    2 i \, q_\psi \, \sin(2\tau)
    +q_\psi^2 \cos^2 (2\tau) 
  \right] P (\tau) = 0 \,,
\end{eqnarray}
where $q_\psi = 2 g_F \Phi /m_\phi$.
As before, we shall use the method of variation of parameters
and write
\begin{eqnarray}
  P (\tau) = y_1(\tau) \, u_1 (\tau) + y_2 (\tau) \, u_2 (\tau) \,,
\end{eqnarray}
where $y_1 = \cos \tau$ and $y_2 = \sin \tau$ are 
the solutions of (\ref{eq:pk}) for the case
$q_\psi = 0$.  Together with the condition (\ref{eq:C_VP})
we obtain the equations for $u_1$ and $u_2$ as
\begin{eqnarray}
  \frac{d}{d\tau}
  \left(
    \begin{array}{c}
      u_1 \\ u_2
    \end{array}
  \right)
  \eqn{=}
  \left[ 
    2 \, q_\psi \, \sqrt{1- \beta_\psi^2} \, \cos (2 \tau )
    - 
    2 i \, q_\psi \, \sin(2\tau)
    +q_\psi^2 \cos^2 (2\tau) 
  \right]
  \nonumber \\
  \eqn{}\times
 \left(
   \begin{array}{c c}
     \sin \tau \, \cos \tau & \sin^2 \tau \\
     - \cos^2 \tau & - \sin \tau \, \cos \tau 
   \end{array}
 \right)
 \left(
   \begin{array}{c}
     u_1 \\ u_2
   \end{array}
 \right) \,.
\end{eqnarray}
By integrating over the period of the $\phi$ oscillation,
the averaged $\overline u_1$ and $\overline u_2$ satisfy
\begin{eqnarray}
  \frac{d}{d\tau}
  \left(
    \begin{array}{c}
      \overline u_1 \\
      \overline u_2
    \end{array}
  \right)
  = - \frac{1}{2}
  \left(
    \begin{array}{c c}
      i q_\psi & q_\psi \sqrt{1- \beta_\psi^2} \\
      q_\psi \sqrt{1-\beta_\psi^2} & - i q_\psi
    \end{array}
  \right)
  \left(
    \begin{array}{c}
      \overline u_1 \\
      \overline u_2
    \end{array}
  \right) \,.
\end{eqnarray}
From the initial conditions (\ref{eq:IC_PH}) at the leading order,
namely, $P(0)=(1+({1-\beta_\psi^2})^{1/2})^{1/2}$
and $dP(0)/d\tau = - i P(0)$, we obtain the solution
\begin{eqnarray}
  \label{eq:TAV_PK}
  P (\tau) =
  P (0)
  \left[ 
    e^{- i \tau} \, \cos \left( \frac{ q_\psi \beta_\psi \tau}{2} \right)
    +
    i \frac{ \sqrt{1- \beta_\psi^2} - 1}{\beta_\psi} \,
    e^{i \tau} \sin \left( \frac{ q_\psi \beta_\psi \tau}{2} \right)
  \right] \,.
\end{eqnarray}
It is then found from (\ref{eq:FPSI}) that the occupation number
for the growing mode is given by
\begin{eqnarray}
  \label{eq:TAV_FPSI}
  f_\psi (t, k_\ast) 
  = 
  \sin^2 \left( 
    \frac{q_\psi \, \beta_\psi \, \tau }{2} \right)
  =
  \sin^2 \left( 
    \frac{g_F \, \Phi \, \beta_\psi }{2} \,  t \right)
  \,.
\end{eqnarray}
Note again that this reproduces the initial behaviour in
Eq.~(\ref{eq:FPSI_ANA}) for $t \ll t_{\rm cr} = 2/(g_F \Phi
\beta_\psi)$, as in the scalar production. 
In Fig.~\ref{fig:F_MODE_gm8_L}, $z_{\rm cr}$ is around 300 when $g_F=10^{-8}$
and $\beta_{\psi}=0.5$.  On the other hand, for $t \gg t_{\rm cr}$, the
occupation number oscillates around $f_\psi = 1/2$ and does not exceed
one, which should be contrast to the scalar production.  This behaviour
reflects the Pauli blocking effect of the produced fermion $\psi$.  As
a result, the number density of $\psi$ stops to grow at $t \simeq
t_{\rm cr}$ even if the coupling $g_F$ is extremely small.

As in the case of scalar production, it can be confirmed that the
solution in Eq.~(\ref{eq:TAV_PK}) satisfies the exact relation
between the mode functions at different times.
For all the mode 
$P(t,k)$ can be written as
\begin{equation}
  P(t , k)
  =
  P(0 , k)\left[P^{(1)}(t , k)-i \, \omega_k(0) P^{(2)}(t , k)\right]\, .
\end{equation}
Here $P^{(1,2)}(t,k)$ correspond to linear independent solutions 
of the equation of motion with the initial conditions
$P^{(1)}(0,k)=\dot{P}^{(2)}(0,k)=1$ and 
$\dot{P}^{(1)}(0,k)=P^{(2)}(0,k)=0$.
These solutions satisfy the same scaling property as Eq.~(\ref{eq:MAP_CHIK}) 
by replacing $\chi_k^{(1,2)}$ into $P^{(1,2)}$ and the occupation number of $\psi$ 
at $t=n T$ is obtained by same manner as
\begin{eqnarray}
  f_\psi(nT,k)
  \eqn{=}
  \frac{k^2}{\omega_k(T)^2}\frac{\sin^2(n \, d)}{\sin^2(d)}
  \left[{\rm Im}P^{(1)}(T,k)\right]^2 \nonumber\\
  \eqn{=} \left[ \frac{\sin (n \, d)}{\sin \left( \left(n-1\right) \, d \right)}\right]^2 f_\psi\bigl( (n-1)T , k \bigr) \,,
  \label{eq:FPSI_NT}
\end{eqnarray}
where $d\equiv \cos^{-1}\left( {\rm Re}P^{(1)}(T)\right)$.  This
equation has already been presented in Ref.~\citen{Greene:1998nh},
however, the estimation of $P^{(1)}(T,k)$ is essential
to obtain $f_\psi(n
T,k)$ by using Eq.~(\ref{eq:FPSI_NT}).  

Now the time averaging method gives us
the analytical expression of $P(t,k_\ast)$ for the growing mode.
Therefore, we can estimate $f_\psi(n T ,k_\ast)$
analytically by taking $P^{(1)}(t,k_\ast)$ and $P^{(2)}(t,k_\ast) $ as
\begin{eqnarray}
  \label{eq:DEC_PK1}
 P^{(1)}(t,k_\ast)\eqn{=}\cos\left(\frac{m_\phi t}{2}\right)\left[
   \cos\left(\frac{g_F \, \Phi \, \beta_\psi}{2} t\right)
   -i \frac{1}{\beta_\psi}\sin\left(\frac{g_F \, \Phi \, \beta_\psi}{2} t\right)
   \right]\nonumber\\
   \eqn{}-\frac{\sqrt{1-\beta_\psi^2}}{\beta_\psi}\sin\left(\frac{m_\phi t}{2}\right) 
   \sin\left(\frac{g_F \, \Phi \, \beta_\psi}{2} t\right) \, ,\\
   \label{eq:DEC_PK2}
P^{(2)}(t , k_\ast) \eqn{=} -\frac{2}{m_\phi}
  \frac{ \sqrt{1-\beta_\psi^2} }{\beta_\psi} \cos\left(\frac{m_\phi t}{2}\right)
  \sin\left(\frac{g_F \, \Phi \, \beta_\psi}{2} t\right)\nonumber\\
  \eqn{} +\frac{2}{m_\phi}\sin\left(\frac{m_\phi t}{2}\right)\left[
   \cos\left(\frac{g_F \, \Phi \, \beta_\psi}{2} t\right)
   +i \frac{1}{\beta_\psi}\sin\left(\frac{g_F \, \Phi \, \beta_\psi}{2} t\right)
   \right]\, .
\end{eqnarray}
Substituting Eq.~(\ref{eq:DEC_PK1}) into Eq.~(\ref{eq:FPSI_NT}),
we obtain the occupation number at $t=n T$ with exact scaling property as
\begin{equation}
  f_\psi(nT,k_\ast) = \sin^2\left(\frac{g_F \, \Phi \, \beta_\psi}{2} n T\right)\,.
\end{equation}
Similar to the scalar production this result is consistent with (\ref{eq:TAV_FPSI}) obtained 
by the time averaging method.

Thus, we conclude that
the results by the time averaging method can not only abstract the
characteristic evolution due to the non-perturbative correction but
also satisfy the exact scaling property in terms of oscillation period
of $\phi$.

\section{Conclusion}
\label{sec:Conc}
We have investigated the particle production from the coherent
oscillation by using the method based on the Bogolyubov
transformation.  For the case when the coupling constants of the
oscillating field are very small, we have obtained the leading
contributions to the distribution functions and the number densities
of the produced particles.

When the amplitude of the oscillation is small ($\Phi \ll
\vev{\phi}$), the leading contributions to the yields are found to be
${\cal O}(g_{S,F}^2)$.  We have presented the exact expressions for
the distribution functions of the produced $\chi$ and $\psi$ at the
${\cal O}(g_{S,F}^2)$ order.  It has been shown that there exists the
growing mode with $k = k_\ast$ if $m_{\chi, \psi} < m_\phi/2$, and its
occupation number increases at the rate $t^2$ after sufficient numbers
of the oscillation.  
The distribution function has a peak at $k \simeq k_\ast$ and the
width of the peak decreases at the rate $1/t$.  As a result, the
number density of produced particles is proportional to $t$.  The
expression for the number density is found to be consistent with the
one obtained by assuming that the coherent oscillation is a correction
of non-relativistic scalar particles and the decay process
is a main source of the particle production.  

We have found that the above perturbative results
fail to describe the exact ones for sufficient late times since the
non-perturbative correction becomes significant even when the coupling
constants are extremely small.  Indeed, the occupation number of
$\chi$ for the growing mode increases exponentially while that of
$\psi$ oscillates around $1/2$.  These distinctive features represent
the statistical properties of the produced particles, \ie, the effects
of the Bose condensation for the scalar production or the Pauli
blocking for the fermion production.  Due to these non-perturbative
effects, the explosive production of $\chi$ happens while the
production of $\psi$ becomes insignificant for late times.  To handle
with these non-perturbative effects we have used the time averaging
method, and have successfully described the evolution of the
occupation number for the growing mode.  This method works well 
because the typical time scale of the evolution is much longer than
the rapid $\phi$ oscillation.  Furthermore, we have shown that
the results obtained by the time averaging method satisfies the
exact scaling properties in Ref.\citen{Mostepanenko:1974}, which also gives
the justification of the use of the time averaging method.

Throughout this analysis, we have neglected the back-reaction effect
of the produced particles in the estimation of the yields.  When the
occupation number of these particles is close to unity, such an effect
should be taken into account.  In addition to this, the inclusion of
the expansion of the universe is also necessary to reveal the
reheating/preheating processes in the inflationary universe.  These
issue will be discussed in elsewhere~\cite{AN}.

\section*{Acknowledgements}
The work of T.A. was partially supported by the Ministry of Education, Science, Sports and Culture, Grant-in-Aid for Scientific Research, 
No.~21540260, and by Niigata University Grant
for Proportion of Project.

\appendix
\section{Derivations of Eqs.~(\ref{eq:NCHI_ANA})
and (\ref{eq:NPSI_ANA})}
\label{sec:AP1}
We show here the derivations of the number densities $n_\chi$ and $n_\psi$ 
at the leading order given in
Eqs.~(\ref{eq:NCHI_ANA}) and (\ref{eq:NPSI_ANA}).  Let us first
consider Eq.~(\ref{eq:NCHI_ANA}).  It is found from
Eqs.~(\ref{eq:F_CHI}) and (\ref{eq:B_K}) that the leading
$\mathcal{O}(g_S^2)$ contribution to $n_\chi$ is given by
\begin{eqnarray}
  n_{\chi}(t)
  \eqn{=}
  \frac{g_S^2 \, \Phi^2 \, m_\chi^2 \, m_{\phi}^2 }{8 \pi^2} \,
  K_\chi (t) \,,
\end{eqnarray}
where
\begin{eqnarray}
  \label{eq:APF_chi}
  &&K_\chi (t) 
  =
  \int_0^tdt_1 \, \sin (m_{\phi}t_1)
  \int_0^tdt_2 \, \sin (m_{\phi}t_2) \,  J_\chi (\Delta t) \,,
  \\
  \label{eq:APG_chi}
  &&J_\chi (\Delta t)
  =
  \int_0^{\infty}dk \, \frac{k^2}{\omega_{\chi}^4} \,
  \cos  ( 2 \omega_{\chi} \Delta t ) \,,
\end{eqnarray}
with $\Delta t = t_2 - t_1$.
The integration in Eq.~(\ref{eq:APG_chi}) can be done as
\begin{eqnarray}
  \label{eq:APG_chi2}
  J_\chi(\Delta t)
  \eqn{=}
  \frac{\pi}{4m_\chi}
  \left[
    1 + 4 \, m_\chi^2 \Delta t^2 - 4 \,m_\chi |\Delta t| 
    \,
    \mbox{${}_1$$F_2$} \left( -\frac{1}{2}; 1, \frac{3}{2};
      -m_\chi^2 \Delta t^2 \right)
  \right] \,,
\end{eqnarray}
where $_1F_2(a;b,c;x)$ is the generalized hypergeometric function.
We then expand $J_\chi$ in terms of $m_\chi$ as
\begin{eqnarray}
    J_\chi (\Delta t)
    =
    \frac{\pi}{m_\chi}
    \left[ \frac{1}{4} + (\Delta t m_\chi)^2
      - \abs{\Delta t} m_\chi 
      \sum_{n=0}^{\infty} \frac{(-1)^{n+1}}{(4n^2 -1)(n!)^2}(\Delta t \, m_\chi)^{2n}  \right] \,.
\end{eqnarray}
Now, we can perform the integrations of $t_1$ and $t_2$ 
in (\ref{eq:APF_chi}) as
\begin{eqnarray}
  K_\chi(t) 
  \eqn{=}
  \frac{\pi}{m_{\phi}^2} \, t \, 
  \left[
    1 - 2 r_\chi^2
    - 2 r_\chi^4
    - 4 r_\chi^6
    -10 r_\chi^8
    + \mathcal{O} (r_\chi^{10})
  \right] + ... 
  \nonumber \\
  \eqn{\simeq} 
  \frac{\pi}{m_{\phi}^2} \, t \, 
  \beta_\chi+ ... 
  \,,
\end{eqnarray}
where $r_\chi = m_\chi/m_\phi$.
Notice that we we have only listed the terms proportional to $t$ 
and dropped off the oscillation terms.  
Finally, we obtain the number density up to
$\mathcal{O}(g_S^2)$ as Eq.~(\ref{eq:NCHI_ANA})
\begin{eqnarray}
    n_{\chi}(t)
    \eqn{=}
    \frac{g_S^2 \, \Phi^2 \, m_\chi^2 \, m_{\phi}^2}{8\pi^2} 
    \, K_\chi(t)
    \simeq
    \frac{g_S^2 \, \Phi^2 \, m_\chi^2 \, \beta_\chi }{8\pi} \,  t \,,
\end{eqnarray}
by neglecting the oscillation terms.  

Next, we turn to consider Eq.~(\ref{eq:NPSI_ANA}).
As in the case of the scalar production, 
the leading ${\cal O}(g_F^2)$ contribution to
the number density of $\psi$ is given by
\begin{eqnarray}
  n_{\psi}(t)
  \eqn{=}
  4 \, \frac{g_F^2 \, \Phi^2 \, m_{\phi}^2 }{8 \pi^2} \,
  K_\psi (t) \,,
\end{eqnarray}
where a prefactor 4 counts the internal degrees of
freedom of $\psi$ and
\begin{eqnarray}
  &&K_\psi (t)
  =
  \int_0^tdt_1 \, \sin (m_{\phi}t_1)
  \int_0^tdt_2 \, \sin (m_{\phi}t_2) \,  
  J_\psi (\Delta t) \,,
  \\
  &&J_\psi(\Delta t)
  =
  \int_0^{\infty}dk \,\frac{ k^4}{\omega_{\psi}^4} \,
  \cos  ( 2 \omega_{\psi} \Delta t ) \,.
\end{eqnarray}
Notice that, comparing with Eq.~(\ref{eq:APG_chi2}),
the integrand of $J_\psi$ has an extra factor $k^2$.

When $m_\psi =0$, we first integrate $t_1$ and $t_2$ in $K_\psi$ and then
the $k$ integration in $J_\psi$ gives
\begin{eqnarray}
  K_\psi (t) 
  \eqn{=}
  \frac{\pi t}{4}
  \left( 1 - \frac{ \sin 2 m_\phi t}{2 m_\phi t} \right) \,.
\end{eqnarray}
On the other hand, when $m_\psi \neq 0$, we first estimate 
$J_\psi (\Delta t)$ as
\begin{eqnarray}
  J_\psi (\Delta t)
  \eqn{=}
  - \frac{ \pi m_\psi}{4} 
  \left[
    3 + 4 m_\psi^2 \Delta t^2
    - 6 m_\psi \abs{\Delta t}
    {}_1F_2 \left( - \frac 12; \frac 32, 2; - m_\psi^2 \Delta t^2 \right)
  \right] 
  \nonumber \\
  \eqn{=}
  - \frac{ \pi m_\psi}{4} 
  \left[
    3 + 4 m_\psi^2 \Delta t^2
    - 6 m_\psi \abs{\Delta t}
    \sum_{n=0}^\infty \frac{(-1)^{n+1}}{(4 n^2 -1) (n + 1) (n !)^2} (\Delta t \, m_\psi)^{2 n}
  \right] 
  \,,~~~
\end{eqnarray}
After the integrations over $t_1$ and $t_2$
we find apart from the oscillation terms
\begin{eqnarray}
  K_\psi (t) = 
  \frac{\pi}{2} t 
  \left[ -3 r_\psi^2 + 3 r_\psi^4 + 4 r_\psi^6 + 3 r_\psi^8 
  + {\cal O}( r_\psi^{10}) \right]+\dots \,,
\end{eqnarray}
where $r_\psi = m_\psi/m_\phi$.  Combining the above two cases, we obtain
\begin{eqnarray}
  K_\psi (t) \eqn{=} \frac{\pi}{4} t 
  \left[ 1 - 6 r_\psi^2 + 6 r_\psi^4 + 4 r_\psi^6 + 6 r_\psi^8 
    + {\cal O}( r_\psi^{10}) \right]+\dots \nonumber\\
  \eqn{\simeq} \frac{\pi}{4} t \beta_\psi^3 +\dots \,,
\end{eqnarray}
which gives Eq.~(\ref{eq:NPSI_ANA})
\begin{eqnarray}
    n_\psi(t) 
    \simeq \frac{g_F^2 \, \beta_\psi^3 \, \Phi^2 \, m_\phi^2}{8 \pi} \, t \,,
\end{eqnarray}
by neglecting the oscillation terms.


\end{document}